\documentclass[pre,reprint,https://v2.overleaf.com/project/5ba006002181f90dc7afddb8, superscriptaddress, nobibnotes]{revtex4-2}
\usepackage{CJK}
\usepackage{amsmath}
\usepackage{amssymb}
\usepackage{graphicx} 
\usepackage{dcolumn}  
\usepackage{bm}       
\usepackage[caption=false]{subfig}
\usepackage{hyperref}
\usepackage{xurl}
\usepackage{float}
\usepackage{xcolor}
\hypersetup{colorlinks=true, pdfstartview=FitV, linkcolor=blue, citecolor=black, plainpages=false, pdfpagelabels=true, urlcolor=blue, breaklinks=true}
\usepackage[all]{hypcap}
\usepackage[normalem]{ulem}
\usepackage{tabu}

\newcommand{\upd}{\text{d}}

\begin{document}

\title{Clonal dynamics of surface-driven growing tissues}

\author{Ruslan I. Mukhamadiarov}
\email{ruslan.mukhamadiarov@lmu.de}
\affiliation{Arnold Sommerfeld Center for Theoretical Physics and Center for NanoScience, Department of Physics, Ludwig-Maximilians-Universität München, Theresienstrasse 37, D-80333 Munich, Germany}
\affiliation{Max Planck Institute for the Physics of Complex Systems, Nöthnitzer Strasse 38, Dresden, D-01138, Germany}
\author{Matteo Ciarchi}
\affiliation{Arnold Sommerfeld Center for Theoretical Physics and Center for NanoScience, Department of Physics, Ludwig-Maximilians-Universität München, Theresienstrasse 37, D-80333 Munich, Germany}
\affiliation{Max Planck Institute for the Physics of Complex Systems, Nöthnitzer Strasse 38, Dresden, D-01138, Germany}
\author{Fabrizio Olmeda}
\affiliation{Max Planck Institute for the Physics of Complex Systems, Nöthnitzer Strasse 38, Dresden, D-01138, Germany}
\author{Steffen Rulands}
\email{rulands@lmu.de}
\affiliation{Arnold Sommerfeld Center for Theoretical Physics and Center for NanoScience, Department of Physics, Ludwig-Maximilians-Universität München, Theresienstrasse 37, D-80333 Munich, Germany}
\affiliation{Max Planck Institute for the Physics of Complex Systems, Nöthnitzer Strasse 38, Dresden, D-01138, Germany}


\begin{abstract}
The self-organization of cells into complex tissues relies on a tight coordination of cell behavior. Identifying the cellular processes driving tissue growth is key to understanding the emergence of tissue forms and devising targeted therapies for aberrant growth, such as in cancer. Inferring the mode of tissue growth, whether it is driven by cells on the surface or cells in the bulk, is possible in cell culture experiments, but difficult in most tissues in living organisms (\emph{in vivo}). Genetic tracing experiments, where a subset of cells is labeled with inheritable markers have become important experimental tools to study cell fate \emph{in vivo}. Here, we show that the mode of tissue growth is reflected in the size distribution of the progeny of marked cells. To this end, we derive the clone-size distributions using analytical calculations in the limit of negligible cell migration and cell death, and we test our predictions with an agent-based stochastic sampling technique. We show that for surface-driven growth the clone-size distribution takes a characteristic power-law form with an exponent determined by fluctuations of the tissue surface. Our results show how the mode of tissue growth can be inferred from genetic tracing experiments.
\end{abstract} 

\maketitle

\section{Introduction}

The self-organization of cells into tissue relies on the coordination of cell proliferation and differentiation in space and time. Broadly, tissue growth can be driven by the spatially homogeneous proliferation of cells (bulk growth). This mode of growth is characteristic of softer tissues like tendroins, arteries, or brain~\cite{shraiman2017}. Alternatively, tissues may grow by the preferential proliferation of cells on the surface, for example, because these cells have access to signaling molecules or vasculature. Surface-driven growth is often found in shells, horns, some bones \cite{shraiman2017}, or tumors, where cells on the tumor surface have better access to nutrients \cite{doi:10.1080/10409238.2019.1611733}. As a specific example of surface-driven growth, in some fish and amphibians the eyecup forms by cell division in the outer part of the eye, the ciliary marginal zone \cite{Marcucci2016}. Understanding whether a given tissue grows by cell proliferation on its surface or in its bulk is important for targeting treatments during aberrant growth, such as cancer, it can form a template for developing synthetic tissues, and for understanding pathological development scenarios. In the example of the eyecup, cell divisions outside of the ciliary marginal zone, in the bulk, lead to the formation of additional blood vessels and scar tissue, and eventually to a disorder called proliferative retinopathy and to a complete loss of the eye's functionality \cite{Heavner2012-gk}. 

The regulation of cell proliferation and the ensuing spatial distribution of proliferation patterns is governed on the one hand by complex biochemical signaling networks and cell-to-cell communication \cite{doi:10.1146/annurev-biophys-070816-033602}. On the other hand, it relies on how microscopic mechanical parameters, such as stresses, translate to a macroscopic scale. The connection between both is not well understood \cite{Lenne2022}, such that an a priori inference of the mode of tissue growth is infeasible from a tissue mechanics perspective  \cite{10.1242/dev.110.1.1, doi:10.1146/annurev-biophys-070816-033602}. Live imaging gives access to spatio-temporally resolved cell kinetics and allows for the estimation of tissue mechanical parameters \cite{10.1242/dev.060103,PhysRevE.76.021910, PhysRevE.77.031907, Rulands2018}. However, live imaging is highly challenging \emph{in vivo} and it is usually limited to specific cases of embryonic development \cite{doi:10.1089/zeb.2008.0562} or to studies of epithelium or other surface tissues \cite{FARHADIFAR20072095,AIGOUY2010773}. 

The recent advent of genetic tracing experiments allows studying cell-fate behavior \emph{in vivo}. In these experiments, a subset of cells is genetically labeled with fluorescent markers or genetic barcodes \cite{Clayton2007}. As cells divide, these labels are passed on to all progeny of a labeled cell, termed a \emph{clone}. The probability density of the sizes of such clones provides indirect information about the history of cell division, differentiation, and cell death events between labeling and the time point of analysis \cite{10.1242/dev.125.1.85, Kretzschmar2012-ij}. For example, the first moment of the clone size distribution, the average clone size, reflects the rate of proliferation and whether both daughter cells remain proliferative or not. The functional form of the clone size distribution reflects how the fate of individual cells is decided \cite{PhysRevE.76.021910, PhysRevE.77.031907, Klein2010-df}, and the presence of mechanical forces \cite{Gibson2006} leading to clone fragmentation and merging\cite{Rulands2018, RULANDS201757, 10.1242/dev.060103}. Therefore, the combination of genetic tracing experiments and tools from statistical physics has become a standard method for unveiling cell-fate behavior \emph{in vivo} \cite{Blanpain2013, Greulich2023, 10.1242/dev.194399, Linus}.

Here, we derive a theory that can help identify the mode of tissue growth from genetic tracing experiments. By studying the stochastic dynamics of clone boundaries and employing ideas from the range expansion process \cite{doi:10.1073/pnas.0710150104,hallatschek2010life}, we show that, for surface-driven tissue growth, the clone size distribution follows a characteristic power-law decay. The decay exponent depends on the roughness of the surface, which in turn is determined by the mechanisms regulating the tissue interface. We confirm our theoretical predictions with Monte Carlo stochastic lattice simulations with forward labeling. 

This paper is organized as follows: In the following section, we introduce the model for surface-driven tissue growth and our mean-field scaling argument for the clone size distribution. We also show that clonal dynamics of surface-growing tissues can be mapped to the first-passage problem of a random walk in the presence of the absorbing boundary. In Sec.~\ref{sec:level3}, we present results for the agent-based lattice simulations of modified Eden cluster growth with label forwarding. We conclude in Sec.~\ref{sec:level4} with a summary and discussion of our main results.

\section{Theoretical results}\label{sec:level2}

\subsection{Model description}

We consider a tissue in $d$ spatial dimensions which are separated from other tissues by a $d-1$ dimensional boundary. At a time $t=0$ a random subset of cells is labeled, and when cells divide, this label is passed on to all progeny of a labeled cell. We are interested in how the size distribution of the number of cells that carry a given label at time $t$ relates to the growth mode of the tissue. To begin our analysis of the clonal dynamics in surface-driven growing tissues, we make a simplifying assumption that cells labeled with the same marker remain spatially segregated. For this to hold true, the rate of cell death needs to be small compared to the rate of cell proliferation, which is generally the case for growing tissues. Moreover, the typical length scale associated with cell migration also needs to be small compared to the spatial extension of clones that result during the time course of the experiment. Under these conditions, cells that share the same marker form spatially-segregated clones such that boundaries that separate clones with different markers are well-defined. Such spatially segregated domains have indeed been observed in experiments of the growing retina of medaka fish \cite{10.7554/eLife.42646}. Under these assumptions, we show that the clone size distribution can be obtained from stochastic and geometric arguments alone without making further assumptions about tissue mechanics.

As the tissue grows, the boundaries of clones are subject to stochastic fluctuations, which originate from the random nature of cell divisions at the tissue surface \cite{doi:10.1073/pnas.0710150104,hallatschek2010life}. In the following, we will first derive expressions for the clone-size distribution in two spatial dimensions and then extend these results to three spatial dimensions. To this end, we will consider the stochastic wandering dynamics of clone boundaries -- an approach that was first applied in the context of the random genetic drift of the range expansion process \cite{doi:10.1073/pnas.0710150104,hallatschek2010life}. 

In two dimensions, the clone boundary dynamics stemming from the stochasticity of cell divisions can be defined by a stochastic process, $X(t)$. The difference in distance between two adjacent clone boundaries, $\Delta X$, has, as the label does not influence proliferation, a vanishing mean, $\langle\Delta X\rangle = 0$, and the time evolution of its variance is described by a wandering exponent $\zeta$,
\begin{equation}\label{eqn:diff}
    \langle (\Delta X)^2\rangle \sim t^{2\zeta} .
\end{equation}
As an example, a wandering exponent $\zeta=1/2$ describes the Brownian motion of the distance between clone boundaries. The presence of tissue surface fluctuations may alter the wandering exponent, e.g., for tissue interfaces that are described by the Kardar-Parisi-Zhang equation, the wandering exponent takes a value $\zeta = 2/3$~\cite{PhysRevLett.56.889, doi:10.1073/pnas.0710150104}.

\begin{figure}[t!]
    \centering
    \includegraphics[width=\columnwidth, trim={0.0cm 0 1.2cm 1cm},clip]{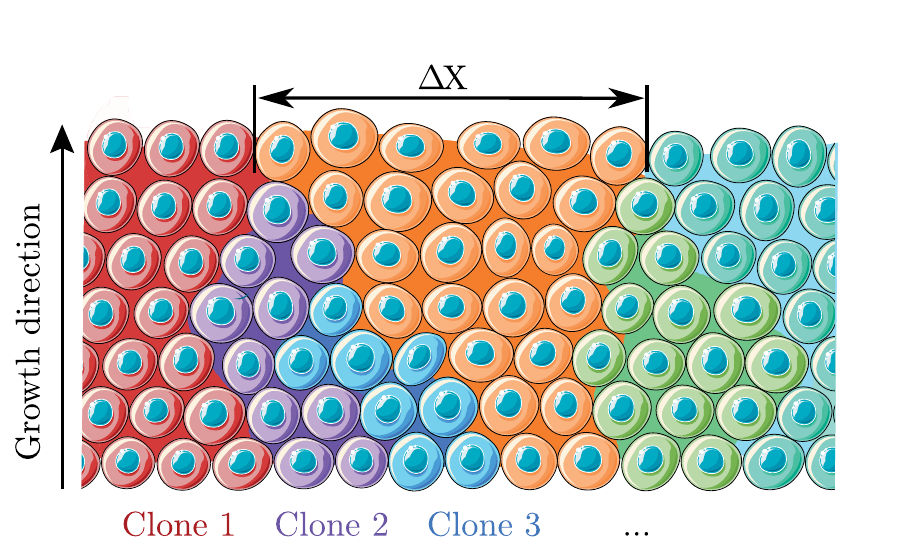}
    \caption{Schematic illustration of expected clonal dynamics in surface-driven growing tissues -- tissues where the cell divisions occur predominantly at tissue's surface. Cells that share the same color belong to the same clone. The arrow indicates the growth direction, and $\Delta X$ indicates the distance between two clone boundaries.}\label{fig:schematics}
\end{figure}

As the tissue grows, adjacent clone boundaries are subject to stochastic coalescence events. Thereby, a clone that is enclosed by two merging boundaries loses its access to the growing tissue surface (see Fig.~\ref{fig:schematics}). As a result of this merging event of the domain boundaries, the number of persisting clones, $N_p$, i.e., the number of clones that have access to the front and that continue to forward their label and grow in size, decreases with time as
\begin{equation}\label{eqn:av_per}
    N_p \sim 1/\sqrt{\langle (\Delta X)^2\rangle} \sim t^{-\zeta}\, .
\end{equation}

\subsection{Mean-field scaling argument}

In order to derive the size of persisting clones, we note that, in contrast to non-growing or bulk-driven tissues, the clonal dynamics in surface-driven tissue growth depend on the proximity of the clone to the surface of the growing tissue. Only clones containing cells at the tissue's surface can continue growing and contributing to the asymptotic shape of the clone-size distribution. If the linear extension of the tissue surface stays constant at a value $L$, we get an approximate expression for the average size of persistent clones, $\langle n_p\rangle$, by dividing the total tissue area by the number of persistent clones at time $t$,
\begin{equation}\label{eqn:av_size}
    \langle n_p(t)\rangle \sim L v\,t / N_p \sim t^{1 + \zeta}\, .
\end{equation}
Here, $v$ is the growth rate of the tissue that we assume to stay constant for a given cell division rate \cite{CYWang_1995, 10.7554/eLife.58945}. 

Asymptotically, the fraction of clones with access to the surface vanishes. Therefore, the clone size distribution, $P(n)$, is well approximated by collecting the sizes of clones that have lost their access to the front, i.e., by collecting the clones that have reached their terminal size. To calculate $P(n)$, we therefore first calculate the number of clones that have lost their access to the moving front in a time interval $\upd t$,
\begin{equation}\label{eqn:N_lost}
    \begin{aligned}
   N_{lost}(t)\upd t =& -\left[ N_p(t + \upd t) -  N_p(t)\right]\upd t \\
   &\sim -(\upd  N_p(t)/\upd t)\,\upd t\\
   &\sim t^{-\zeta - 1}\upd t\, .
    \end{aligned}
\end{equation}
Then, using the mean-field argument that all persisting clones grow with the same average rate $n(t) = \langle n_p(t)\rangle$ in Eq.~\eqref{eqn:av_size}, we obtain the clone-size distribution for surface-driven growing tissues,
\begin{equation}\label{eqn:pdf_algebraic}
    \begin{aligned}
        P(n)\upd n &= N_{lost} (\upd t/\upd n) \upd n \sim n^{-1} n^{-\zeta/(1 + \zeta)}\upd n\\
            &= n^{-(1+2\zeta)/(1+\zeta)} \upd n\, .
    \end{aligned}
\end{equation}
The clonal size distribution has a unique, previously unobserved, power-law form, with an exponent that only depends on the wandering exponent $\zeta$ that describes the stochastic motion of clone boundaries. This result is in contrast to log-normal distribution and exponential distributions observed for bulk-driven growing tissues and in homeostatic tissues, respectively \cite{Rulands2018, PhysRevE.76.021910, PhysRevE.77.031907}.

In three-dimensional tissues, clone boundaries are defined by stochastic surfaces. For a given clone, consider a slice along the direction of the growth. Within this slice, we expect the distance between the clone boundaries, $\langle (\Delta X)^2\rangle$ to scale like $t^{2\zeta}$. In the absence of anisotropies, this scaling holds for all $d-1$ directions perpendicular to the growth direction. We now consider the number of cells that share the same marker in a given slice perpendicular to the growth direction, $A$. Its deviation from the average, $\Delta A$, fluctuates as 
\begin{equation}\label{eqn:area_fluct}
    \langle (\Delta A)^2 \rangle \sim  (\langle\Delta X^2 \rangle)^2 \sim t^{4\zeta}\, .
\end{equation}
If the number of cells in a given slice remains constant, the number of clones that have access to the surface decreases as (cf. Eq.~\eqref{eqn:av_per})
\begin{equation}\label{eqn:Np_3d}
    N_p \sim 1/\sqrt{\langle (\Delta A)^2 \rangle} \sim t^{-2\zeta}\, .
\end{equation}
For growth with a constant growth rate $v$, the average size of persistent clones increases as
\begin{equation}
    \langle n_p(t)\rangle \sim L^2 v\,t / N_p \sim t^{1+2\zeta}\, . \label{eqn:aver_3d}
\end{equation}
Finally, utilizing the same line of argument as for two-dimensional tissues, the clone size distribution in $d=3$ reads
\begin{equation}\label{eqn:PDF_3d}
    P(n) \upd n \sim n^{-(1+4\zeta)/(1+2\zeta)} \upd n\, .
\end{equation}

Taken together, these scaling arguments predict that the clone size distributions follow characteristic power laws. The associated exponents depend on the spatial dimension and the wandering exponent of clone boundaries, which is again influenced by the roughness of the tissue surface. For flat surfaces, where $\zeta=1/2$, the clone size distribution decays with an exponent of $4/3$ for planar tissues and $3/2$  for volumnar tissues. For a large class of fluctuating surfaces belonging to the Kardar-Parisi-Zang universality class ($\zeta=2/3$), the clone size distribution decays with exponents $7/5=1.4$ and $11/7\approx 1.57$, in $d=2$ and $d=3$ respectively.

We derived these results in the limit of negligible curvature. For curved tissue surfaces, clone boundary coalescence halts asymptotically if the mean squared displacement of clone boundaries increases slower than the expansion of the tissue surface~\cite{hallatschek2010life, Aif2022,10.7554/eLife.42646}. Therefore, the results are strictly valid if $2\, \zeta > d-1$ for surfaces with constant curvature. Even if this is not the case, our results are applicable if the linear extension of clones, $\Delta X$, is much smaller than the length scale associated with the curvature. This is generally the case not too long after labeling. Since genetic tracing experiments typically trace clones over several rounds of cell divisions, we expect our results to be broadly applicable.

\begin{figure*}[t!]
    \centering
    \hspace{0.0cm}
    \includegraphics[width=2\columnwidth, trim={0.0cm 0 0cm 0cm},clip]{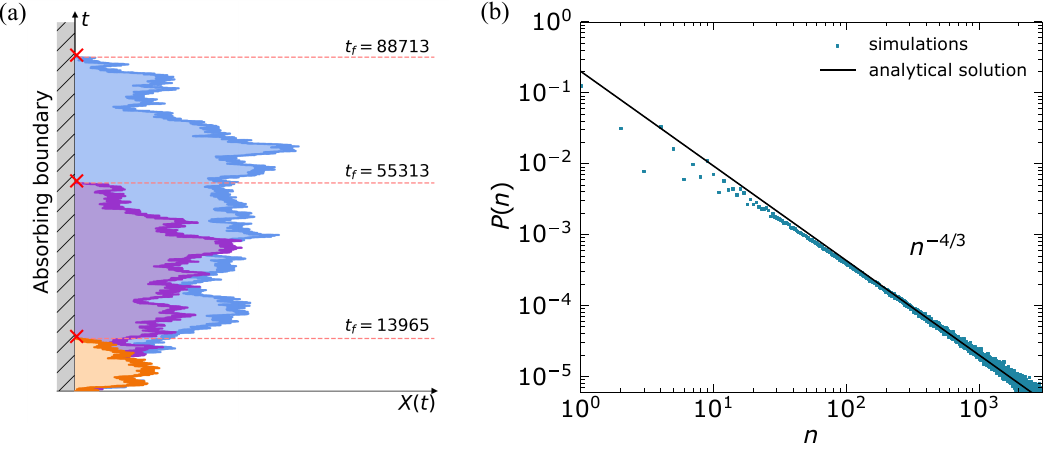}
    \caption{(a) Sample trajectories of one-dimensional Brownian walker in the presence of an absorbing boundary at $X=0$. The enclosed areas can be obtained by evaluating the integral of first-passage Brownian functional $\mathcal{A} = \int_0^{t_f}X(t)dt$, where $t_f$ is the first time the process $X(t)$ crosses the origin. 
    (b) Probability distribution $P(n)$ that area $\mathcal{A}$ takes a specified value $n$, when the Brownian motion begins at $X=0$: comparison of the analytical prediction by \cite{Majumdar_2020} with our Monte-Carlo simulations.}\label{fig:RW_absorbing}
\end{figure*}

\subsection{Analogy with the first-passage problem}

The power-law form of clonal size distribution can be obtained by associating coalescence events of clone domain boundaries that follow $\langle (\Delta X)^2\rangle \sim t$ with the first passage events of a Brownian walker that hit the origin (see Fig.~\ref{fig:RW_absorbing} (a)). Specifically, if the distance between two clone boundaries performs a random walk as the tissue front advances, then, since the position of the tissue interface $h$ depends explicitly on time (e.g., $h\sim vt$), the final clone size can be associated with the area $\int X(t) dt$ that the random walker would cover before it hits the absorbing boundary at the origin. As such, the size distribution of clones that reached their final size is equal to the size distribution of the areas subtended by a random-walk trajectory 
\begin{equation}
   \{X(0), X(t_1),..., X(t_f)\}\, ,
\end{equation}
where $t_f$ is the first-passage time of the stochastic process $X(t)$ hitting the absorbing boundary. 
\newline

For diffusive motion, in the continuum limit, the first-passage Brownian functional can be written as
\begin{equation}
\mathcal{A} = \int_0^{t_f}X(t)dt\, .
\end{equation}
It has been evaluated analytically in \cite{Majumdar_2020}, where, for large areas, the authors demonstrated that the probability density function $P(\mathcal{A})$ follows a power law 
\begin{equation}
    P(\mathcal{A}) \sim \mathcal{A}^{-4/3}.
\end{equation}
This result is in agreement with what we obtained with the mean-field argument for a Brownian case $\zeta=1/2$. We also confirmed this result by running Monte Carlo simulations for the one-dimensional random walk in the presence of an absorbing boundary at $X=0$, for which we computed the areas $A=\sum_{t=0}^{t_f}X_t$ covered by the Brownian walker before it hits the absorbing boundary (Fig.~\ref{fig:RW_absorbing} (b)). For anomalous diffusion, $\zeta\neq 1/2$, exact solutions for the first-passage functional $\mathcal{A}$ are not known apart from specific stochastic processes \cite{Mattia_PRE2023}.\\

\section{Agent-based lattice simulations}\label{sec:level3}

To test the validity of our analytical predictions, we performed numerical simulations of surface-driven growth in $d=2$ and $d=3$. For these simulations to verify the predicted power-law exponents they need to generate clones spanning orders of magnitude in size. Simulations of such extent are impossible when considering tissue mechanics and the details of biochemical processes underlying cell fate regulation. However, for surface-driven growth, if the rate of cell motility and loss are significantly smaller than the rate of cell division, the clone-size distribution is expected to depend only on the wandering and coalescence statistics of clone domain boundaries, and not on the underlying tissue mechanics or regulatory biochemical signaling network. Therefore, we use computationally efficient lattice simulations that capture the stochastic dynamics of clone domain boundaries and their relation to the clone-size distribution without necessarily being accurate microscopic representations of the tissue mechanics and regulatory processes. 

Specifically, we employ a modified version of the Eden cluster growth model, which is a minimal agent-based model that produces surface-driven cluster growth \cite{murray1961two, CYWang_1995}. In addition to the diffusion-limited branching process $A\xrightarrow{\lambda} A + A$ that increases the size of clusters by 1 with a rate $\lambda$, we randomly label agents $A$ at the beginning of the simulation and allow them to pass their label upon replication. To produce and sample clone statistics, we employ Monte Carlo simulations of the described diffusion-limited birth process with label forwarding on two- and three-dimensional lattices.
All of our simulations are initialized with a fully-occupied line ($d=2$) or plane ($d=3$) at $x=0$, while the rest of the lattice is empty. Initially, each agent is endowed with a unique label. We update the system state using the Monte Carlo random sequential updating scheme. For a randomly chosen agent, we select at random an empty nearest neighbor lattice site and generate a new agent with the same label with a rate $\lambda$. 

\begin{figure*}[t!]
    \centering
    \subfloat[\label{fig:fig3ab}]{\includegraphics[width=2\columnwidth,trim={0cm 0 0.0cm 0.0cm},clip]{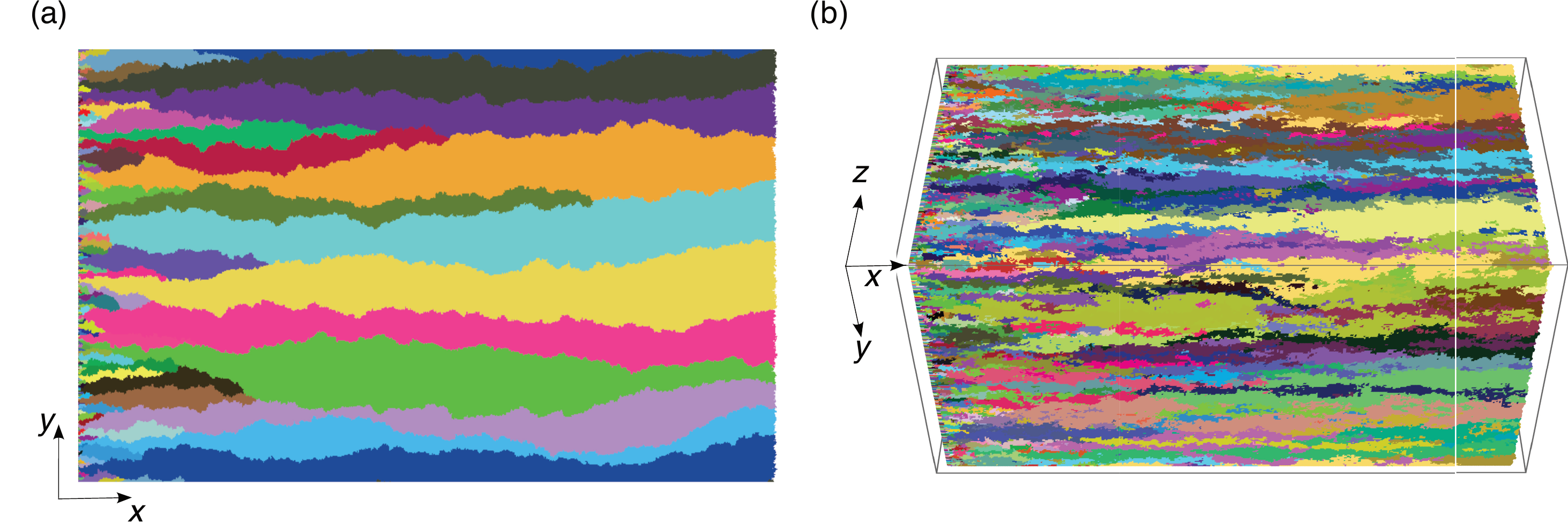}}
    \vfill
    \vspace{-0.5cm}
    \subfloat[\label{fig:fig3cd}]{\includegraphics[width=2\columnwidth,trim={0cm 0.6cm 0.0cm 0.0cm},clip]{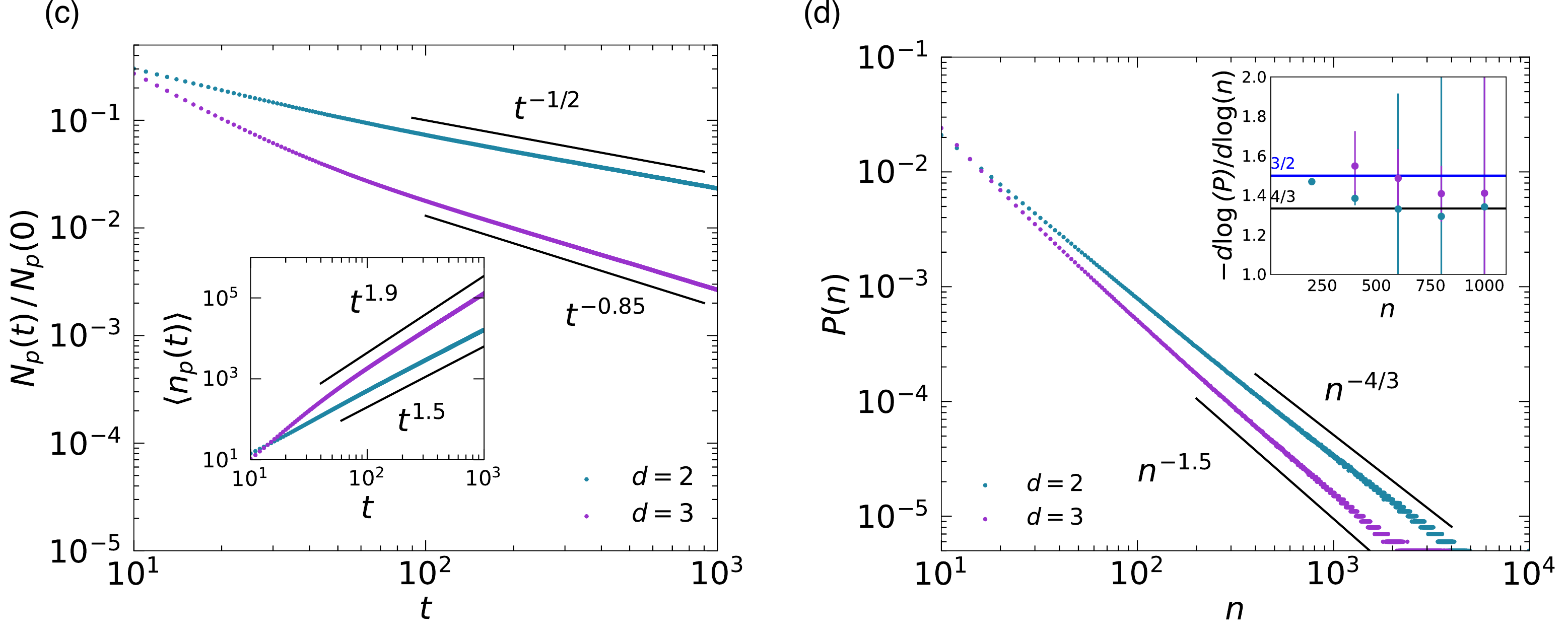}}
    \caption{Snapshots of Monte Carlo lattice simulation of the birth process with label forwarding for (a) two-dimensional $1000\times500$ lattice and (b) three-dimensional $400\times200\times200$ lattice. In both cases, the simulation begins at $X=0$ with only the first perpendicular layer being filled with agents, each having a unique label that they can forward when they reproduce. The snapshots are taken right before the front reaches the other end of the lattice $X=L_x$. For both (a) and (b) we have kept periodic boundary conditions in directions perpendicular to the growth. (c) Time evolution of the number of persisting clones $N_p(t)$ divided by its initial value $N_p(0)$ and the average size of persisting clones $\langle n_p(t)\rangle$ (inlay). Both quantities are obtained from two- and three-dimensional Monte Carlo lattice simulations of a simple birth process with label forwarding, and time is measured in Monte Carlo steps. (d) The clonal size distribution and its local decay exponent. The inlay shows the local exponent. Error bars depict $\pm$ standard deviation. The data in (c) was obtained from simulations on $1000\times500$ and $1000\times100\times100$ latices and were averaged over $10^4$ independent realizations. The data for the clonal size distribution in (d) was obtained from simulations on $500\times200$ and $100\times50\times50$ lattices and were averaged over $10^6$ independent simulation runs.}\label{fig:simulations_flat}
\end{figure*}

\begin{figure*}[t!]
    \centering
    \subfloat[\label{fig:fig4ab}]{\includegraphics[width=2\columnwidth,trim={0cm 0 0.0cm 0.0cm},clip]{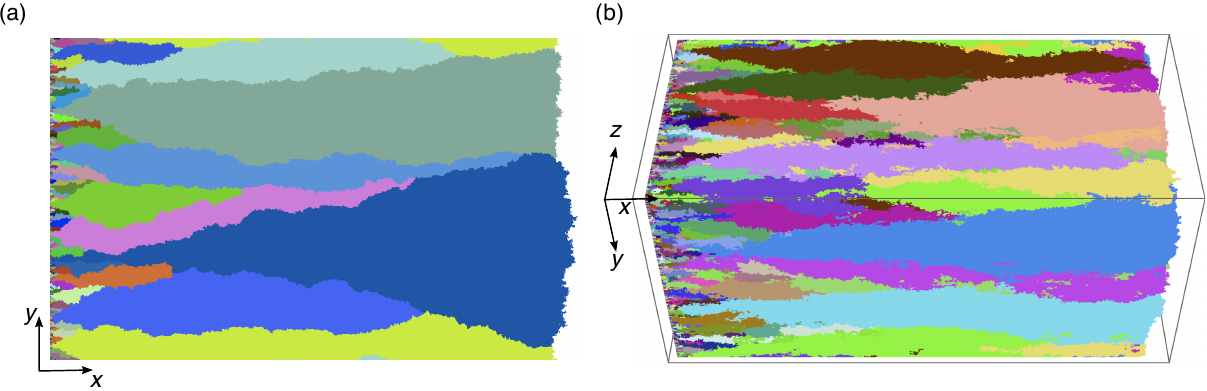}}
    \vfill
    \subfloat[\label{fig:fig4cd}]{\includegraphics[width=2\columnwidth,trim={0cm 0.6cm 0.0cm 0.0cm},clip]{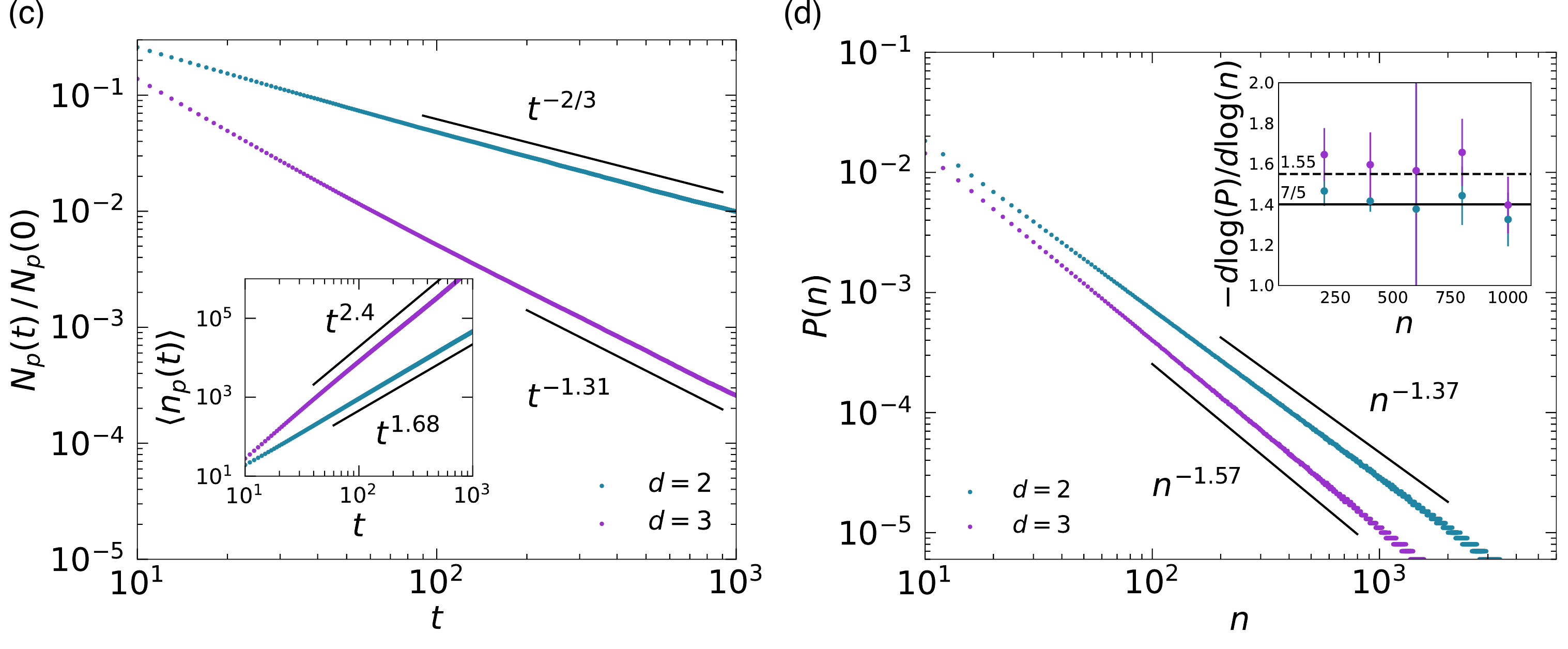}}
    \caption{Snapshots of Monte Carlo lattice simulation of the birth process with constant rate $\lambda=\text{const.}$ with label forwarding for (a) a two-dimensional $1000\times500$ lattice and (b) a three-dimensional $400\times200\times200$ lattice. In both cases, the simulation begins at $X=0$ with only the first perpendicular layer being filled with agents, each with a unique label they can forward when they reproduce. The snapshots are taken right before the front reaches the other end of the lattice $X=L_x$. For both (a) and (b), we have kept periodic boundary conditions in directions perpendicular to the growth. (c) Time evolution of the number of persisting clones $N_p(t)$ divided by its initial value $N_p(0)$ and the average size of persisting clones $\langle n_p(t)\rangle$ (inlay). Both quantities are obtained from two- and three-dimensional Monte Carlo lattice simulations of a simple birth process with label forwarding with $\lambda=\text{const.}$, and time is measured in Monte Carlo steps. (d) The clonal size distribution and its local decay exponent, obtained from simulations with a rough KPZ front. The inlay shows the local exponent. Error bars depict $\pm$ standard deviation. The data in (c) was obtained from simulations on $1000\times500$ and $1000\times100\times100$ latices and were averaged over $10^4$ independent realizations. The data for the clonal size distribution in (d) was obtained from simulations on $500\times200$ and $100\times50\times50$ lattices and were averaged over $10^6$ independent simulation runs.}\label{fig:simulations_KPZ}
\end{figure*}

First, we simulated systems where clones have boundaries that follow Brownian dynamics, i.e., $\zeta=1/2$ in Eq.~\eqref{eqn:diff} and Eq.~\eqref{eqn:area_fluct}, respectively.
This situation is realized for tissues with sharp and smooth interfaces, where surface fluctuations can be neglected. 
We achieve that in our simulations by choosing a space-dependent growth rate, $\lambda = (1- \tanh\left[\alpha(x - x_0) \right])/2$, where the coefficient $\alpha$ sets the surface sharpness, $x_0 = vt$ determines the surface position, and $v$ sets the growth velocity. Choosing $\lambda$ to have a sigmoidal functional form that varies only along the growth direction prohibits the development of surface fluctuations in directions perpendicular to the growth. As shown in Fig.~\ref{fig:simulations_flat} (a)-(b) for $d=2$ and $d=3$ cluster growth, respectively, the cluster interface stays flat at all times for this choice of $\lambda$, and clone boundaries perform a random walk. We then collect the number of different labels $N_p$ that have access to the front and measure the size of these clones to compute $\langle n_p\rangle$. To obtain the clonal size distribution, we collected the sizes of the clones that have lost their access to the advancing surface. 

As shown in Fig.~\ref{fig:simulations_flat} (c)-(d), our simulations reflect the predictions made by the scaling arguments given above with some slight deviations for $d=3$ cluster growth. Specifically, all our results for two-dimensional growth follow the predicted values after the early-time transient. We attribute this early-time transient to relaxation from the initial flat, sharp interface with zero width to the steady-state interface with a finite $1/\alpha$ width. In three spatial dimensions, as shown in Fig.~\ref{fig:simulations_flat} (c), the number of persistent clones $N_p$ and the average size of persistent clones $\langle n_p\rangle$ deviate slightly from our mean-field analysis. It is plausible that the slower decay in the number of persistent clones $N_p$ for $d=3$ stems from the fragmentation and merging clones, which can occur in $d=3$ and is not considered in the mean-field theory. Nevertheless, if we substitute measured power laws for $N_p^{\text{sim}} \sim t^{-0.85}$ and $\langle n_p^{\text{sim}}\rangle \sim t^{1.9}$ into Eq.~\eqref{eqn:N_lost} and Eq.~\eqref{eqn:pdf_algebraic}, the value for the clonal size distribution exponent that is predicted by the mean-field argument agrees well with our simulation data. This fact supports the connections that we have established when deriving the mean-field expression for the clone size distribution.

If the tissue interface is rough in the sense that the interface gradients are pronounced, tissue surface height fluctuations can no longer be neglected. As has been shown in \cite{PhysRevLett.74.4325, doi:10.1073/pnas.0710150104}, the boundaries between clones receive an additional drift contribution that comes from a surface tilt, rendering their dynamics superdiffusive, i.e., $\zeta>1/2$. In situations when the tissue interface evolution is described by the KPZ equation \cite{PhysRevLett.56.889}, which is often the case for rough interfaces, the expression for the wandering exponent becomes $\zeta = 1 + (\chi - 1)/z$ \cite{doi:10.1073/pnas.0710150104}. In this expression, the dynamical exponent $z$ describes how the characteristic linear extension $L_x$ of the surface height fluctuation grows with time $L_x \sim t^{1/z}$, while the roughness exponent $\chi$ determines the scaling ratio of the fluctuation height to fluctuation's linear extension $L_z \sim L_x^{\chi}$. For a one-dimensional tissue interface ($d=2$, $d_{\text{surf}}=1$) whose dynamics is described by the KPZ equation, the exponents that characterize the scaling of surface fluctuations would belong to $1+1$ KPZ universality class and take $\chi = 1/2$,  $z=3/2$ values \cite{PhysRevLett.56.889}. Consequently, the wandering exponent will be equal to $\zeta = 2/3$ and, according to our mean-field argument, $P\sim n^{-7/5}$ in $d=2$. For three-dimensional tissues with two-dimensional KPZ interface, $\chi\approx 0.382$ and $z=1.618$ \cite{GOMESFILHO2021104435}, which results in $\zeta = 0.618$ and $P\sim n^{-1.55}$.

We first confirm that the Eden cluster growth with constant agent division rate $\lambda$ produces a rough traveling front with KPZ fluctuations \cite{meakin1998fractals, PhysRevLett.87.238303, PhysRevLett.113.180602}. We do that in our $d=2$ and $d=3$ simulations by computing the width of the interface height fluctuations $W(L,t)$ at the front of the growing Eden cluster:
\begin{equation}
    W(L,t) = \sqrt{\frac{1}{L}\int_0^{L} dx\, (h(x,t) - \langle h(t)\rangle)^2}\, ,
\end{equation}
and then, employing the Family-Viscek scaling $W(L,t) = L^{-\chi}\,\overline{W}\left(t/L^z\right)$, we confirm that the surface height fluctuations are characterized by the KPZ scaling exponents.

After that, we use the same simulation algorithm to generate clones with different sizes, i.e., we begin with only the line or plane at $X=0$ being occupied with agents that have unique labels, and we allow agents to pass their labels as they reproduce. Once the tissue interface reaches the other end of the simulation box $X=L_x$, we stop the simulation and collect the sizes of the clusters that share the same label. As shown in Fig.~\ref{fig:simulations_KPZ} (a)-(b), in comparison to simulations with a flat front, pronounced surface gradients make clones lose their access to the surface at a faster rate. Measuring the mean-squared displacement of the clone domain boundaries from $d=2$ simulations, we confirm that $\langle (\Delta X)^2\rangle \sim t^{4/3}$ and that the number of the persisting clones drops as $N_p(t)\sim t^{-2/3}$ (Fig.~\ref{fig:simulations_KPZ} (c)). Similarly to the flat interface scenario, the clonal size distribution for simulation with KPZ surface dynamics follows a power-law decay with exponents that come very close to our mean-field predictions, as shown in Fig.~\ref{fig:simulations_KPZ} (d). For $d=3$ simulations, we again observe slight deviations for $N_p$ and $\langle n_p(t)\rangle$ quantities from our mean-field predictions, which we attribute to possible branching and merging of the clones.

\section{Conclusion}\label{sec:level4}

In summary, we have studied the dynamics of clones for both $d=2$ and $d=3$ surface-driven growing tissues. We found that the clone-size distribution takes a power-law form with exponents depending on the tissue dimension and the nature of fluctuations in the surface. The power laws in the clone size distribution translate to associated power laws in the time evolution of the average clone size and the number of clones with access to the surface. These results suggest how the mode of tissue growth can be identified using genetic tracing experiments. While genetic tracing experiments using fluorescent markers typically do not produce a sufficiently high number of clones to confidently identify such power laws, recently developed technologies using genetic barcodes produce millions of unique clones in tissue \cite{Merino2019} and can, therefore, be used to infer the mode of tissue growth as well as distinguish different kinds of surface fluctuations.

Throughout this work, we have been assuming that the tissue growth by cell divisions at its surface would lead to the formation of clonal sectors, i.e., the cells that share the same label would stay grouped up, and the domain walls that separate different clones would be clearly defined. While we have excluded cell migration and cell turnover from our analysis in order to be able to use this clone domain-wall dynamics approach, both of these processes are essential for tissue growth and remodeling and cannot be excluded from consideration entirely \cite{doi:10.1146/annurev.bioeng.7.060804.100340, Trepat2009, doi:10.1073/pnas.1219937110}. As a potential direction for future research, it would be interesting to build a model that does not rely on the clonal domain wall description and would allow label mixing by including cell death and migration processes. Finally, the model of surface-driven tissue growth that we considered in this work leaves out the tissue mechanics and the effects that may come from it. For example, it would be interesting to consider a situation where cells that have lost access to the tissue interface and are still not far from the surface may continue dividing and regain access to the tissue interface once again.

\begin{acknowledgments}
We thank Martin Lenz, Ricard Alert, and Frank J\"ulicher for valuable discussions. This project has received funding from the European Research Council (ERC) under the European Union’s Horizon 2020 research and innovation program (grant agreement no. 950349). 
\end{acknowledgments}

\bibliographystyle{apsrev4-1}
\bibliography{references}

\end{document}